%% file: jhuevent.tex
\documentclass[10pt,conference]{IEEEtran}
\usepackage{graphicx,color,subfigure,colordvi}
\usepackage{epsfig}	
\usepackage{textcomp}
\usepackage{verbatim,url}
\usepackage{citesort}
\usepackage{amsmath}

\begin{document}
\title{The Perils of Detecting Measurement Faults in Environmental
Monitoring Networks}

\author{Jayant Gupchup$^\ast$ \quad Abhishek Sharma$^+$ \quad Andreas Terzis$^\ast$ \quad Randal Burns$^\ast$ \quad Alex Szalay$^\ddagger$ \\ 
Johns Hopkins University, Computer Science Department$^\ast$ \\ Johns
Hopkins University, Physics and Astronomy Department$^\ddagger$ \\
University of Southern California, Computer Science Department$^+$ }

\maketitle

\input{abstract2}

\input{intro2}

\input{relwork}

\input{method}

\input{results2}

\input{imply}

%\input{ack}

\begin{small}
\bibliographystyle{IEEEtran}
% \bibliography{sensors}

\end{small}

\end{document}

%% file: abstract2.tex
\begin{abstract}

Scientists deploy environmental monitoring networks to discover
previously unobservable phenomena and quantify subtle spatial and
temporal differences in the physical quantities they measure. Our
experience, shared by others, has shown that measurements gathered by
such networks are perturbed by sensor faults. In response, multiple
fault detection techniques have been proposed in the
literature. However, in this paper we argue that these techniques may
mis-classify events (\emph{e.g.} rain events for soil moisture
measurements) as faults, potentially discarding the most interesting
measurements. We support this argument by applying two commonly used
fault detection techniques on data collected from a soil monitoring
network. Our results show that in this case, up to 45\% of the event
measurements are misclassified as faults.  Furthermore, tuning the
fault detection algorithms to avoid event misclassification, causes
them to miss the majority of actual faults.  In addition to exposing
the tension between fault and event detection, our findings motivate
the need to develop novel fault detection mechanisms which incorporate
knowledge of the underlying events and are customized to the sensing
modality they monitor.
\end{abstract}

%% file: intro2.tex
\section{Introduction}
\label{sec:intro}

% General about environmental monitoring

Wireless sensor networks have been used in a number of environmental
monitoring applications~\cite{SSC+06,SWC+07,TPS+05}, offering
scientists the ability to observe physical phenomena in spatial and
temporal granularities not previously possible. In turn, these
observations reveal previously unknown physical phenomena and
subtle variations (\emph{e.g.} micro-climates) that scientists could
not previously measure.

% Fault detection

Alas, environmental monitoring networks introduce their own set of
problems: results from early deployments have shown that sensor faults
occur occasionally, causing faulty data to be recorded and
collected~\cite{TPS+05,ALJL+06,FFCENS06}. The underlying cause of
these faults include incorrect hardware and software design,
malfunctioning transducers and low battery levels. Irrespective of
their origin, faults need to be detected so the network does not
consume its resources in delivering corrupted measurements and these
measurements do not pollute the experiment. Given the importance of
this problem, a number of \emph{fault detection} techniques have been
proposed in the literature~(\emph{e.g.,} \cite{SECON07,AML05} among
others). While each technique uses a different statistical method to
detect faults, they all rely on the assumption that faulty data are
inherently \emph{different} from so-called
\emph{normal} data.

% Event can be misclassified as faults

In this paper we argue that the blanket assumption that all
measurements which do not conform to some notion of normalcy are due
to faults and thus should be discarded, is a precarious one. 
One of the goals of environmental monitoring networks are 
to detect rare and subtle \emph{events}. We buttress
this argument by employing two fault detection techniques, initially
proposed in~\cite{SECON07}, to detect faults in a dataset collected
from a soil monitoring network we deployed. We then measure how many
events (in this context rainfall instances) were classified as
faults. Our results show that these techniques can misclassify up to
45\% of the events as faults. Moreover, tuning the techniques'
parameters such that events are no longer misclassified, leads to a
large number of false negatives, that is failing to detect actual
faults.

% Future 

In addition to identifying and quantifying the danger of
misclassifying events as faults using specific fault detection
algorithms, we provide a list of directions for developing novel fault
detection algorithms that are sensitive to events. Specifically, we
stress the importance of leveraging the \emph{signatures} of events,
as reflected by different modalities, in reducing the number of
misclassifications. We observe that the onset of an
event can be indistinguishable from a fault. Furthermore, different sensors
register delayed versions of the same underlying events, provide
another argument for temporarily storing collected measurements before
relaying them to the back-end.

This paper has five sections. In the section that follows we review
related work in the area of fault and event detection in wireless
sensor networks. Section~\ref{sec:sol} summarizes the two fault
detection techniques, originally presented in~\cite{SECON07}, we use
in this study and describes the types of faults they are designed to
detect. In Section~\ref{sec:res} we present the results of applying
these algorithms to data gathered from a soil monitoring network and
quantify the percentage of events that are misclassified as
faults. Finally, we close in Section~\ref{sec:imply} with a discussion
about the requirements for future fault detection algorithms.

\subsection{Events in Environmental Monitoring Networks}

We present our work using data from a soil monitoring network we
deployed at the Jug Bay wetlands sanctuary. This sanctuary is located
along the Patuxent river in Maryland and serves as the habitat for a
variety of turtle species, including the Eastern Box turtle. These
turtles are of scientific interest because their sex is not determined
by sex genes but by the incubation temperature. It has been shown in
the lab that a difference of two degrees centigrade is enough to
produce male instead of female offsprings. On the other hand, the in
vivo conditions of box turtle nests have not been observed in the
wild. Considering environmental conditions in turtle nests are
currently unknown, correlating rare events with nest conditions could
reveal valuable information.
The network we deployed in Spring 2007 continuously monitored
the conditions of three turtle nests until the eggs hatched in
September of the same year. We use Tmote Sky motes~\cite{PSC05},
coupled with ECH$_2$0 EC-5 soil moisture sensors from Decagon and
custom soil temperature sensors. We also measure the temperature
inside each mote's enclosure using the mote's on-board temperature
sensor. We term this reading \emph{box temperature} to differentiate
it from soil temperature.

\begin{figure}[t]
 \centering
  \epsfig{file=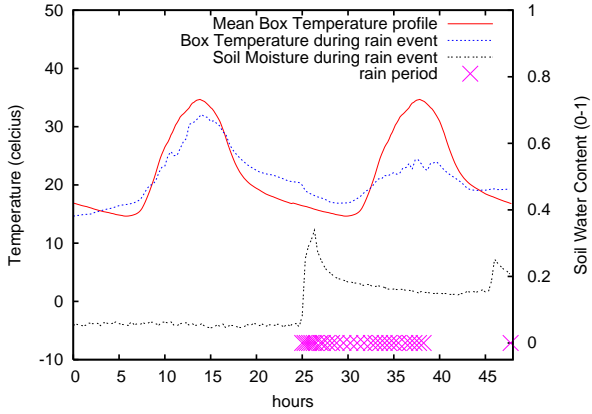, width=0.9\columnwidth}
  \caption {Reaction of box temperature and soil moisture modalities
  to a rain event shown on the x-axis. Soil
  water pressure is measured as the ratio of water volume to the total
  soil volume. Also shown is the normal box temperature
  profile, generated by averaging the box temperature
  at the same time of day during all non-event days.} 
  \label{fig:sig}
\end{figure}

Figure~\ref{fig:sig} presents an example of how box temperature and
soil moisture register a rainfall event. The duration and magnitude of
these events are recorded by a weather station co-located with the
soil monitoring network. The figure shows a non-event day followed by
a day with considerable rain. One can notice that box temperature
during the second day clearly differs from the normal diurnal
temperature pattern. The reaction of soil moisture is distinctly
different---the event's onset causes a sudden increase in the value
recorded by the soil moisture sensor followed by a period of gradual
drying of the soil and corresponding decrease in soil moisture. The
magnitude of the increase and the duration of the decay period are
controlled by the amount and the duration of the rainfall.

%
% AT: It is very hard to see the green line on BW printers. Please change
%

%% file: relwork.tex
\section{Related work}
\label{sec:related}

Fault characterization and detection has received significant
attention in the sensor network community, starting with the work of
Koushhanfar et al. who proposed a cross validation procedure to detect
generalized sensor faults in real time~\cite{SENS03}. More recently,
Ramanathan et al. provided an account of the types and the underlying
causes of sensor faults they encountered in three soil sensor
deployments~\cite{FFCENS06}. Partially motivated by these findings,
Sharma et al. provided a taxonomy of sensor faults and proposed
multiple approaches to detect these faults in real and simulated
datasets~\cite{SECON07}. In this paper, we focus
on understanding how existing fault detection mechanisms perform in
datasets that contain events which deviate from the norm.

%
%Ni et al. provide a
%comprehensive breakdown of the different types of faults and the touch
%on some common reasons underlying these faults~\cite{CENSTSN}.  Their
%contribution lies in recognizing that some faults have information
%value, which should not be thrown away, and, the role played by
%environmental features in modeling data that enables the detection of
%faults.
%
%
%
%
%Deshpande ~\cite{AGM+04} propose a model-based approach to deal with
%irregularities and unreliability in the data.  
%

Abadi et al. introduced a declarative approach for detecting sensor
events~\cite{AML05}. Specifically, they proposed distributing and
storing ``event predicates'' on a network's sensor nodes. The nodes
then compute in-network joins of the collected data and notify the
user when one of the described event predicates are satisfied. We are
interested in understanding whether events can be misclassified as
faults using fault detection techniques proposed in the literature.

%% file: method.tex
\section{Methodology}
\label{sec:sol}

We describe the two fault detection techniques we use and present the
faults they are designed to detect.

\subsection{Types of Faults}

We focus on two types of sensor faults that have been experimentally
observed by a number of environmental monitoring networks. Using the
terminology coined by Sharma et al. in~\cite{SECON07}, we consider
SHORT and NOISE faults.  SHORT faults are characterized by a drastic
difference between the current and the previous sensor measurement. On
the other hand, a NOISE fault is characterized by a period during
which the data samples exhibit larger than normal variations. Sharma
et al. also defined the CONSTANT fault type, in which case the
standard deviation of the collected samples is (almost) zero.  Instead
of defining a third category, we expand the definition of NOISE faults
to include sets of measurements whose standard deviation is
significantly \emph{higher or lower} compared to the overall standard
deviation.

\subsection{Fault detection techniques}
\label{sec:method}

In order to detect the fault types described above and to study the
prevalence of event misclassifications, we implement the
heuristic-based and estimation-based techniques presented
in~\cite{SECON07}.

The heuristic-based techniques consist of the SHORT rule and the NOISE
rule. In the SHORT rule, whenever the absolute difference between the
current and the last measurement is larger than $\delta$, the current
measurement is classified as a fault. The appropriate value of
$\delta$ is obtained from leveraging domain knowledge.  The NOISE rule
declares a fault whenever the standard deviation ($\sigma_{sample}$)
of a set of $N$ successive measurements exceeds a
threshold. Specifically, if $\sigma_{sample}$ is not within
$\sigma_{train} \pm \sigma_{allow}$, we consider all $N$ samples as
faulty. We compute $\sigma_{train}$ by dividing the training data
into sets of $N$ consecutive samples and compute the standard
deviation for each of these sets. We then plot the histogram of all
standard deviation values and set $\sigma_{train}$ to be the mean
value of the histogram. Furthermore, $\sigma_{allow}$ is set as an
integer multiple of the standard deviation of the histogram. In
Section~\ref{sec:res}, we present the effect of varying
$\sigma_{allow}$ on the misclassification error. Finally, we found
empirically that setting the number of samples, $N$, to the equivalent
of a 6-hour time window gave the best results.

The estimation-based technique is an application of linear
least-square estimation (LLSE)~\cite{LLSE} and leverages any
correlations between the measurements collected by spatially
distributed sensors.  Let $i$ and $j$ be two sensors whose
measurements $s_i(t)$ and $s_j(t)$ are correlated. We assume that the
correlation can be represented by a linear model and thereby the
estimate of $s_i(t)$ based on $s_j(t)$ can be written as:

$$\hat{s}_{ij}(t) = \beta_{0,j} + \beta_{1,j}*s_j(t)$$ 

\noindent 
Equivalently, in matrix notation, 
$$\mathbf{S_{ij}} = \mathbf{\tilde{S}_j}\cdot \mathbf{\beta}$$ 
where $\mathbf{S_{ij}}$ is the
vector of $\hat{s}_{ij}$ estimates, $\mathbf{\tilde{S}_j} = [\mathbf{1} \mid
\mathbf{S_j}]$ where $\mathbf{S_j}$ is the vector of $s_j$ measurements,
and $\beta = [\beta_{0,j} \thickspace \beta_{1,j}]^T$.  Using the LLSE formulation, we
set $\beta$ to $
(\mathbf{\tilde{S}_j}^T\mathbf{\tilde{S}_j})^{-1}\mathbf{\tilde{S}_j}^T\mathbf{S_{ij}}$
using measurements from a training set. 

Then, the estimation error is $\epsilon_{ij}(t) =
\hat{s}_{ij}(t) - s_i(t)$. If $\epsilon_{ij}(t)$ is
greater than a threshold, $T_{ij}$, we consider $s_i(t)$ as faulty.
The threshold $T_{ij}$ is set such that $p\%$ of the estimation errors
are below $T_{ij}$ when the model is applied to the training set.  In
practice, we compute the threshold $T_{ij}$ for each of $i$'s
$k$ neighbors and declare the reading $s_i(t)$ as faulty if more than
$q$ neighbors have $\epsilon_{ij} > T_{ij}$.

%% file: results2.tex
\section{Evaluation}
\label{sec:res}

%We begin by establishing metrics to study the misclassification of
%events as faults. We describe our data and the process of injecting
%faults into our test data, creating an evaluation set which contains
%events and faults. Last but not least, we describe our
%results and provide some intuition on their interpretation.

\subsection {Evaluation Metrics}
\label{sec:metric}

In order to study the misclassification of events as faults, we need
to establish appropriate metrics. The misclassification metric we use
has two variants depending on the method under evaluation. First, the
SHORT-rule and the LLSE methods classify individual sensor readings as
faulty and therefore the misclassification error $\mu$ can be defined
as $$
\mu = 
\frac{\textrm{number of event measurements tagged as
faults}}{\textrm{total number of event measurements}}  
$$ 
In this case, an \emph{event measurement} is a sensor reading
(\emph{e.g.} box temperature) during an event (rainfall). 

On the other hand, the NOISE-rule method declares sets of $N$
successive samples as faulty and therefore the metric must account for
the misclassification duration. Let us say that the $i$-th event spans
$E_i$ samples and let $F_i$ be the number of successive samples that
are declared as faulty. Then, all sets of samples $D_i$ within $F_i$
that overlap with $E_i$ contribute to the misclassification error. One
can then obtain the total misclassification error by summing over all
misclassification instances: $$
\mu = \frac {\sum_{i} D_i}{\sum_{i} E_i}
$$
	
Having established a misclassification metric, we need a metric to
study the efficacy of the fault detection method itself. The reason is
that one can set a method's parameters to minimize the number of
misclassifications. Doing so however, might cause the method to fail
to detect the actual faults. We use the false negative ratio, defined
as the fraction of faults that were not detected by the method to the
total number of faults, for this purpose.

\begin{figure}[t]
 \centering
  \epsfig{file=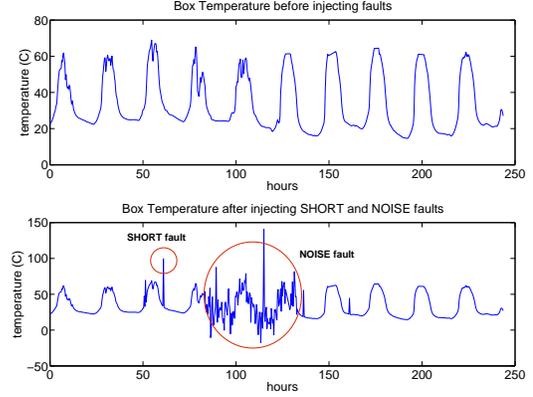, width=0.9\columnwidth} \\
  \caption {Box temperature test data after injecting SHORT and NOISE
  faults.}  
  \label{fig:BTfaults}
\end{figure}

\begin{figure*}[t]
	\centering
	\setcounter{subfigure}{0}
	\subfigure[Soil Moisture SHORT-rule ]{
		\label{sfig:sm_short}\includegraphics[width=0.43\linewidth]{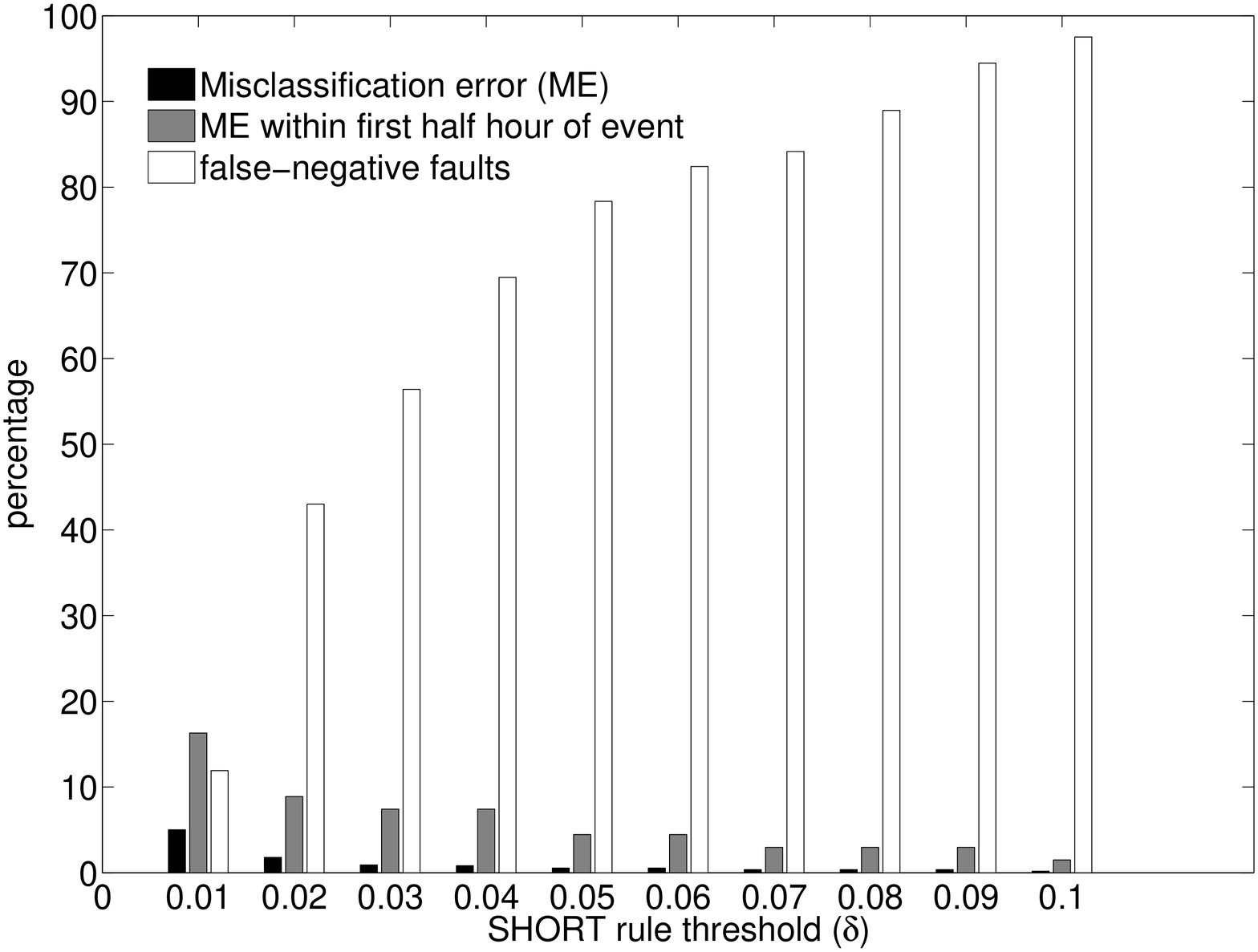}
	}\,
	\setcounter{subfigure}{1}
	\subfigure[Box Temperature SHORT-rule]{
		\label{sfig:bt_short}\includegraphics[width=0.43\linewidth]{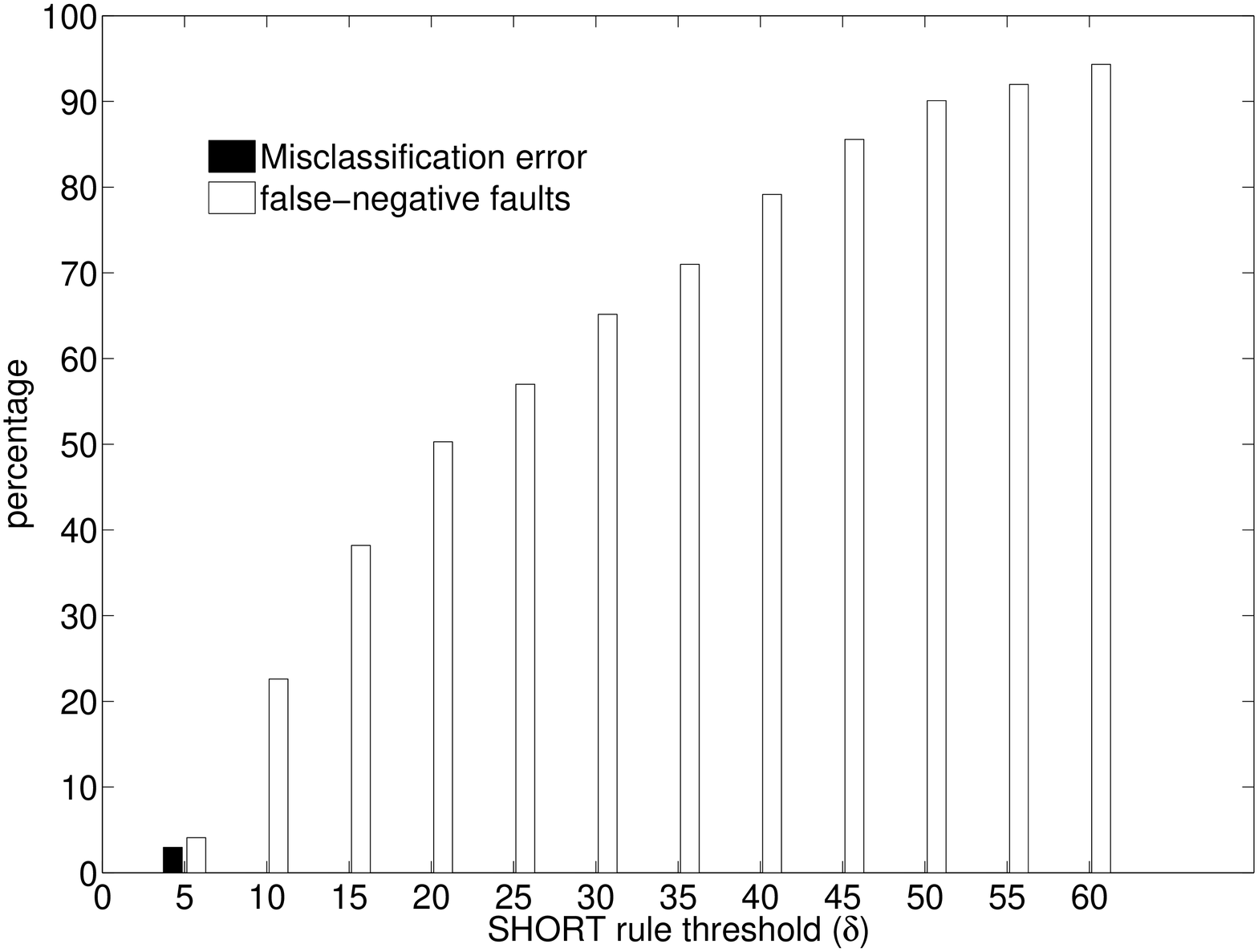}
	}\\
	\setcounter{subfigure}{2}
	\subfigure[Soil Moisture NOISE-rule]{
		\label{sfig:sm_noise}\includegraphics[width=0.43\linewidth]{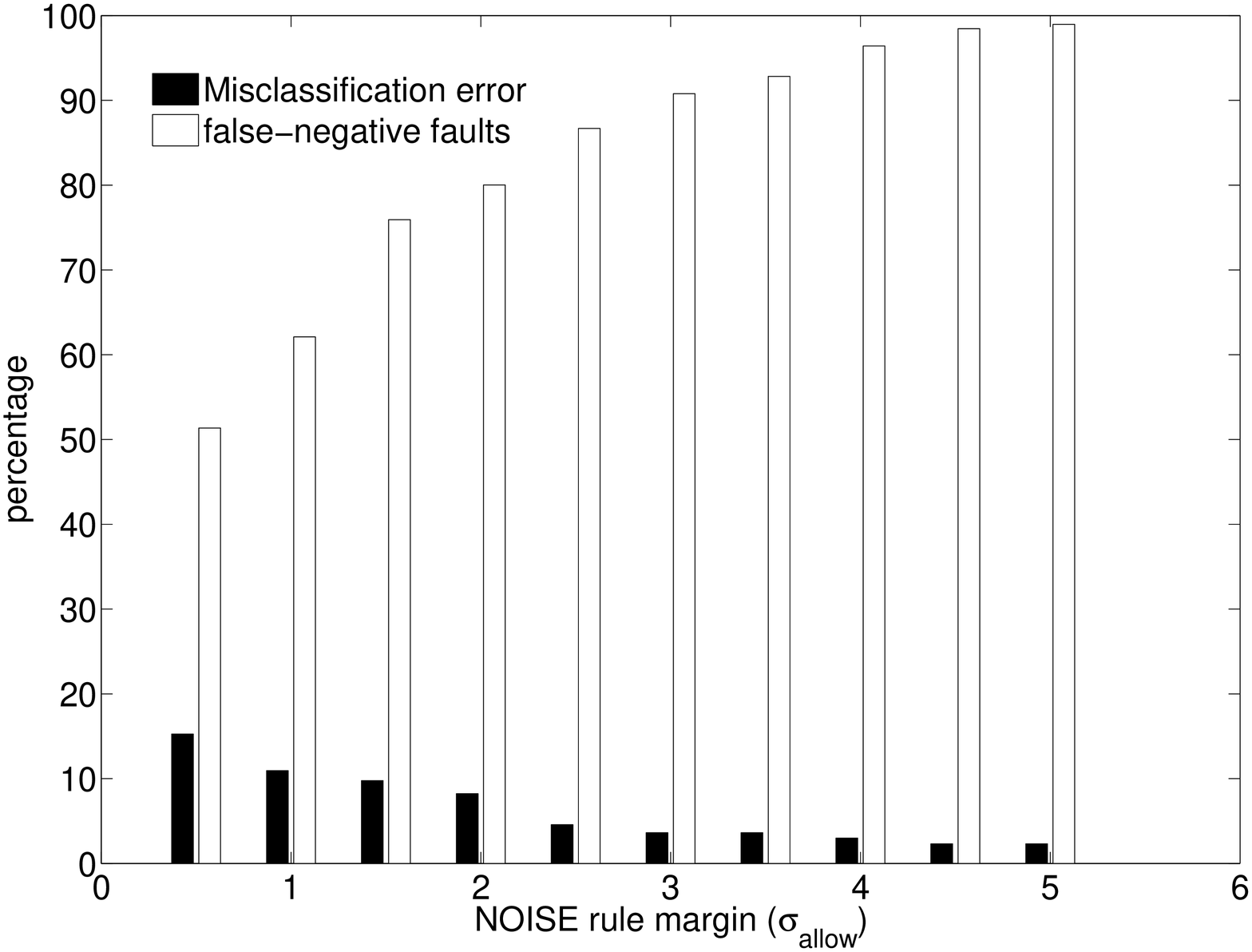}
	}\,
	\setcounter{subfigure}{3}
	\subfigure[Box Temperature NOISE-rule]{
		\label{sfig:bt_noise}\includegraphics[width=0.43\linewidth]{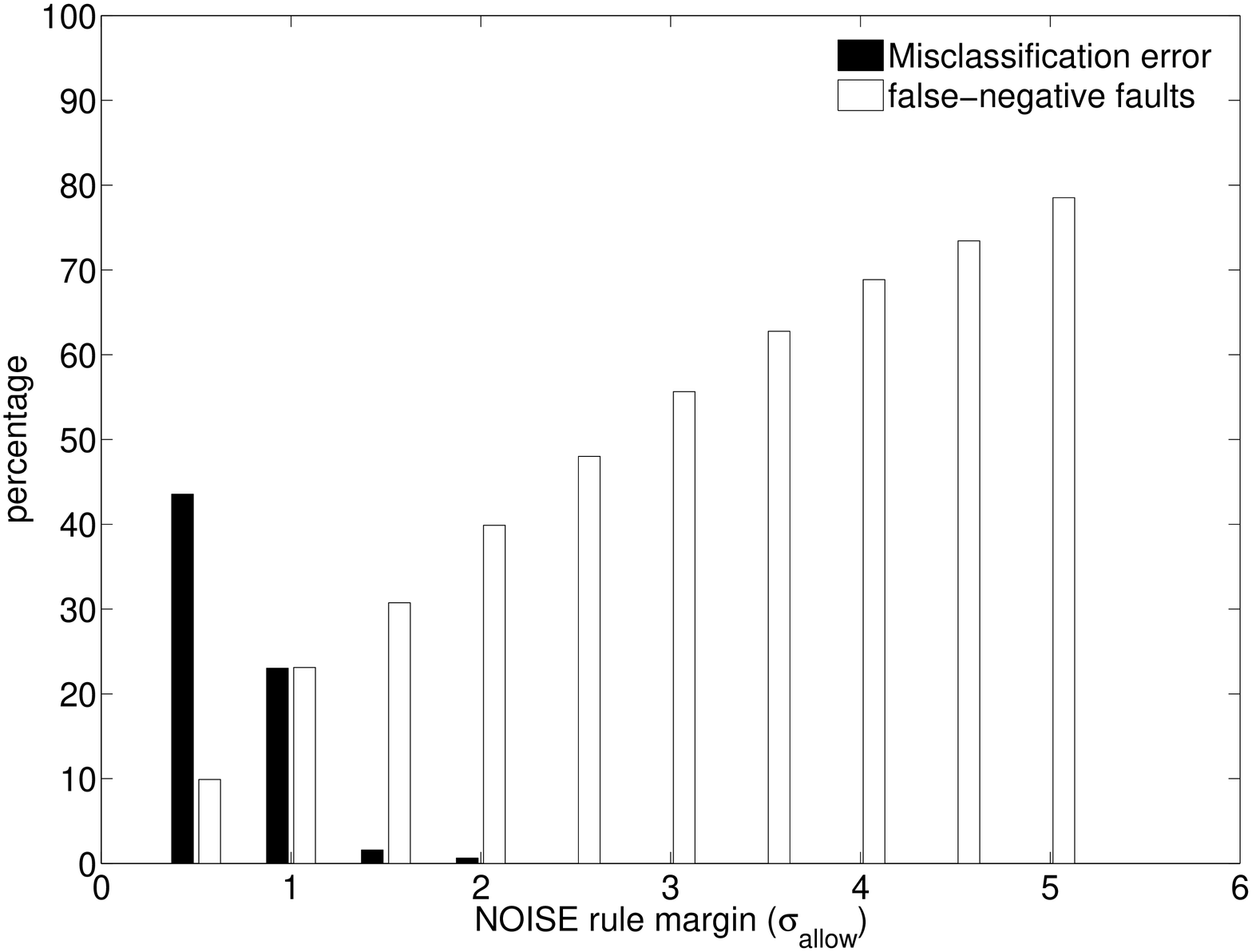}
	}\\

	\caption{Misclassification error and fault negative
	ratios for soil moisture and box temperature 
	using the SHORT and NOISE rules.  Sub-figure (a) also shows the
	misclassification error during the first half hour of rain events.}
\end{figure*}

\subsection{Data}

We apply the fault detection techniques presented in
Section~\ref{sec:method} to the data obtained from the Jug bay turtle
monitoring sensing network ~\cite{AGU07}. Specifically, we use the box
temperature and soil water content (\emph{i.e.}, soil moisture)
modalities collected by three motes at the deployment site. The raw
data series consists of measurements taken at ten minute intervals,
but we use a smoother version by calculating the average of every two
sensor readings.

Approximately five months of data was collected from the Jug bay
sensor deployment, from 2007/06/22 to 2007/11/27.  We use one month of
data from each of the sensors for training purposes and the rest as
test data.  The training data was thoroughly cleaned using a median
filter.  Moreover, we visualized the data and manually removed any
faulty readings to make the training set devoid of faults. The set of
events that occurred during the deployment period is gathered from a
weather station located approximately 700 meters away from the
monitoring site, which records precipitation data at 15-minute
intervals~\cite{JBPR}.  Twenty one major events occurred during the
measurement period, spanning a total of 9,480 rainfall minutes (158
hours).

\subsection{Fault Injection}

As we mentioned in Section~\ref{sec:metric}, we are also interested in
the percentage of real faults that the detection algorithms
miss. However, in order to calculate this ratio we need to know which
measurements correspond to actual faults. Considering that we do not
know which actual sensor readings are faulty, we resort to
artificially injecting SHORT and NOISE faults. To do so, we use the
procedure outlined by Sharma et al.~\cite{SECON07}.

To inject a SHORT fault, a sample $v_i$ is picked at random and is
replaced by the value $\hat{v_i} = v_i + f*v_i$.  SHORT faults with
intensities $f = \{0.5,1,2\}$ and $f = \{0.1,0.2,0.5\}$ were injected
in the test set for box temperature and soil moisture respectively.
To inject a NOISE fault, a set of $W$ successive samples is randomly
chosen and random values drawn from the distribution $\sim N(0,
\sigma^2)$ are added to the test set.  NOISE faults causing an increase
of $0.5\times, 1.5 \times,$ and $3 \times$ in standard deviation
($\sigma$) were injected in the box temperature and soil moisture test
data sets.  The fraction of SHORT faults in the data was set at
$1.5\%$.  We inject NOISE faults consisting of 144 and 360 consecutive
samples, such that the total number of NOISE faults samples in the
data is $6.5\%$. Note that SHORT faults are ephemeral and thus are
much more in number compared to the NOISE faults which last for longer
periods of time. Figure~\ref{fig:BTfaults} provides an example of the
artificially injected fault data for box temperature.

%we set
%he value of $W$ to have $|W|/N$ equal to 3/2 for both box
%temperature and soil moisture. %

% JG :
% I didnt quite follow your explanation of |W| 
% basically NOISE faults that span 144 & 360 readings are 
% inserted. What i was trying to get at earlier was that
% 3/5 of the NOISE faults come from |W=144| and the
% other 2/5 comes from the NOISE faults with |W=360|
%

\subsection{Results}

In order to study the effect of SHORT faults on misclassification, we
evaluate the misclassification error and false negative fraction for
the SHORT rule as a function of increasing
$\delta$. Figure~\ref{sfig:sm_short} and Figure~\ref{sfig:bt_short}
show the results on soil moisture and box temperature respectively. As
one would expect, SHORT faults have a higher impact on soil moisture
misclassification compared to box temperature because rain events
generate measurement spikes that can be mis-interpreted as faults. For
this reason, we find that a significant proportion of the
misclassifications occur in the first half hour of the event period,
jeopardizing the most valuable part of the event data. Even though the
misclassification error decreases as $\delta$ increases, one still
observes considerable misclassification errors during the events'
first half hour. It is clear that this loss can be mitigated by
buffering suspicious data, and leveraging the soil moisture event
signature to discriminate them from faults. The SHORT rule works well
for box temperature data since box temperature does not show an
equivalent leading edge behavior during an event's onset.

Next, we study the misclassification error for the NOISE rule as a
function of $\sigma_{allow}$. Figures~\ref{sfig:sm_noise} and
\ref{sfig:bt_noise} show the performance of the NOISE
rule. The persistent misclassification error for soil moisture across
all $\sigma_{allow}$ values can be explained by the observation that
soil moisture does not show any variation unless when it spikes in
reaction to rain events. Therefore, avoiding misclassifications
requires large $\sigma_{allow}$ values which in turn leads to missing
most actual faults.

On the other hand, as Figure~\ref{sfig:bt_noise} indicates, increasing
$\sigma_{allow}$ does lead to a sharp decrease in misclassification
error for box temperature.  Moreover, we found that most of the
misclassifications are caused due to the lower side of the rule
($\sigma_{train} - \sigma_{allow}$), which is not surprising as box
temperature is known to drop before, during and after a rain event
(cf. Fig.~\ref{fig:sig}). At the same time, increasing
$\sigma_{allow}$ has the undesired side effect of increasing the ratio
of false negatives by threefold.

\begin{figure}[t]
 \centering
  \epsfig{file=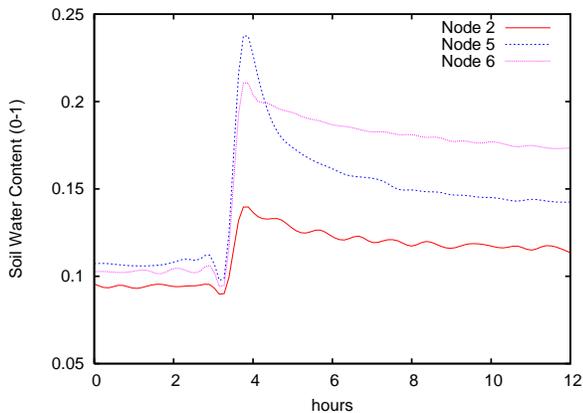, width=0.9\columnwidth} \\
  \caption {Reactions of different soil moisture
  sensors to a rain event.}  
 \label{fig:reaction}
\end{figure}

For the LLSE method, we set the confidence interval, $p$, to 95\% and
the neighbor-opinion, $q$, to two. Table~\ref{tab:llse} presents our
findings for the two sensing modalities. The higher misclassification
error associated with soil moisture can be attributed to the
observation that soil moisture sensors react differently depending on
their location, due to soil's heterogeneity. Figure~\ref{fig:reaction}
presents an example of phenomenon, by plotting the reaction of three
soil moisture sensors to a rain event. Node 5 and Node 6 tend to move
together, whereas the reaction of Node 2 is much lower in magnitude.

%
% AT: How can this be physically explained given the relative
%locations of 2,5, and 6?
% 

\begin{table} [t]
\caption{Misclassifications and false negatives for the LLSE method} 
\centering
\begin{tabular}{|c|c|c|}
\hline
Modality  &  Misclassification & False \\
 & error & Negatives \\
\hline 
Box Temperature & 0.3\% & 77.19\% \\
\hline 
Soil Moisture & 46.3\% & 50.03\% \\
\hline
\end{tabular}

\label{tab:llse}
\end{table}

%% file: imply.tex
\section{Discussion}
\label{sec:imply}

The results from Section~\ref{sec:res} indicate that some fault
detection techniques are susceptible to misclassifying events as
faults. While more sophisticated techniques might be able to reduce
the percentage of misclassifications, the point we want to raise is
that we need a new acid test for \emph{all} fault detection
techniques. This test will evaluate their performance in the presence
of events. Doing so is crucial, because environmental monitoring
networks are in many cases deployed for the purpose of detecting rare
and subtle events that deviate from the norm.

As we noted earlier, events have characteristic signatures that are
specific to each sensing modality.  For example, a rain event causes a
sudden increase in the value of measured soil moisture followed by a
period of gradual decay. The decay rate is a function of multiple
factors, including the soil type, amount of rain, and duration of
rain. Furthermore, the \emph{onset} of a rain event is
indistinguishable from a SHORT fault. This observation suggests that
motes should not prematurely characterize measurements as faults but
should rather buffer enough data points to be able to compare
\emph{suspicious} measurements against event signatures. The same
argument applies to comparing measurements across different motes,
because these motes might be registering the same event with different
time lags. 

While different from the 'baseline' signal, we conjecture that, as the
rain example implies, events follow distinct and common patterns that
can be identified and exploited to reduce misclassifications. As part
of our previous work, we used a Principal Component Analysis (PCA)
based technique to identify the most significant characteristics of the
baseline signal (\emph{i.e.} the daily, seasonal
cycles)~\cite{DSI07}. We believe that a similar methodology can be
used to discover the common characteristics of event signals. This
information can be then encoded and used to differentiate events from
true faults.

%Looking at the problem from the other side, one can ask, what are the
%effects of event detection techniques on fault detection.  We propose
%~\cite{DSI07} a PCA based event detection technique that leverages the
%signatures of these events to classify events. In our experience, some
%faults were misclassified as events. For the purpose of our specific
%application, false positives hurt us far less compared to false
%negatives (missing out on detecting events).  Thus, one needs to
%recognize the duality of events and faults and keep this in mind while
%making application specific design decisions. In practical situations,
%this mis-classification might be inescapable, but making use of
%domain-specific features will definitely reduce the problem to more
%acceptable levels.

%Finally, a question that comes up in both event detection and fault
%detection is: ``does one size fit all?''  As we have stressed
%throughout this paper, different modalities have very different
%characteristics. We also saw that different modalities are subject to
%varying vulnerabilities in detecting faults (and events). The natural
%question to ask is: what is the best way to deal with this
%seamlessly. Does one need to consider different methods tuned to the
%modality, or is there a good way to combine these differences and come
%up with one technique that fits the budget and constraints of a
%typical sensor network.

%% file: jhuevent.bbl
\begin{thebibliography}{10}
\providecommand{\url}[1]{#1}
\csname url@rmstyle\endcsname
\providecommand{\newblock}{\relax}
\providecommand{\bibinfo}[2]{#2}
\providecommand\BIBentrySTDinterwordspacing{\spaceskip=0pt\relax}
\providecommand\BIBentryALTinterwordstretchfactor{4}
\providecommand\BIBentryALTinterwordspacing{\spaceskip=\fontdimen2\font plus
\BIBentryALTinterwordstretchfactor\fontdimen3\font minus
  \fontdimen4\font\relax}
\providecommand\BIBforeignlanguage[2]{{%
\expandafter\ifx\csname l@#1\endcsname\relax
\typeout{** WARNING: IEEEtran.bst: No hyphenation pattern has been}%
\typeout{** loaded for the language `#1'. Using the pattern for}%
\typeout{** the default language instead.}%
\else
\language=\csname l@#1\endcsname
\fi
#2}}

\bibitem{SSC+06}
R.~Mus\u{a}loiu-E., A.~Terzis, K.~Szlavecz, A.~Szalay, J.~Cogan, and J.~Gray,
  ``{Life Under your Feet: A Wireless Sensor Network for Soil Ecology},'' in
  \emph{Proceedings of the 3$^{rd}$ EmNets Workshop}, May 2006.

\bibitem{SWC+07}
L.~Selavo, A.~Wood, Q.~Cao, T.~Sookoor, H.~Liu, A.~Srinivasan, Y.~Wu, W.~Kang,
  J.~Stankovic, D.~Young, and J.~Porter, ``{LUSTER: Wireless Sensor Network for
  Environmental Research},'' in \emph{Proceedings of the 5$^{th}$ ACM Sensys
  Conference}, Nov. 2007.

\bibitem{TPS+05}
G.~Tolle, J.~Polastre, R.~Szewczyk, N.~Turner, K.~Tu, P.~Buonadonna,
  S.~Burgess, D.~Gay, W.~Hong, T.~Dawson, and D.~Culler, ``{A Macroscope in the
  Redwoods},'' in \emph{Proceedings of the 3$^{rd}$ ACM SenSys Conference},
  Nov. 2005.

\bibitem{ALJL+06}
G.~Werner-Allen, K.~Lorincz, J.~Johnson, J.~Lees, and M.~Welsh, ``{Fidelity and
  Yield in a Volcano Monitoring Sensor Network},'' in \emph{Proceedings of the
  7$^{th}$ USENIX Symposium on Operating Systems Design and Implementation
  (OSDI)}, 2006.

\bibitem{FFCENS06}
N.~Ramanathan, T.~Schoellhammer, D.~Estrin, M.~Hansen, T.~Harmon, E.~Kohler,
  and M.~Srivastava, ``The final frontier: Embedding networked sensors in the
  soil,'' UCLA, Center for Embedded Networked Computing, Tech. Rep. CENS-TR-68,
  November 2006.

\bibitem{SECON07}
A.~Sharma, L.~Golubchik, and R.~Govindan, ``On the prevalence of sensor faults
  in real world deployments,'' in \emph{IEEE Conference on Sensor, Mesh and Ad
  Hoc Communications and Networks (SECON)}, 2007.

\bibitem{AML05}
D.~J. Abadi, S.~Madden, and W.~Lindner, ``{REED: Robust, Efficient Filtering
  and Event Detection in Sensor Networks},'' in \emph{Proceedings of the
  31$^{st}$ VLDB Conference}, 2005.

\bibitem{PSC05}
J.~Polastre, R.~Szewczyk, and D.~Culler, ``{Telos: Enabling Ultra-Low Power
  Wireless Research},'' in \emph{Proceedings of the Fourth International
  Conference on Information Processing in Sensor Networks: Special track on
  Platform Tools and Design Methods for Network Embedded Sensors (IPSN/SPOTS)},
  Apr. 2005.

\bibitem{SENS03}
F.~Koushanfar, M.~Potkonjak, and A.~Sangiovanni-Vincentelli, ``On-line fault
  detection of sensor measurements,'' \emph{Proceedings of IEEE Sensors},
  vol.~2, pp. 974--979, Oct. 2003.

\bibitem{LLSE}
R.~Hutchison, \emph{{Linear Least-Square Estimation}}.\hskip 1em plus 0.5em
  minus 0.4em\relax Stroudsburg, PA, 1977.

\bibitem{AGU07}
K.~Szlavecz, A.~Terzis, R.~Musaloiu-E., C.-J. Liang, J.~Cogan, A.~Szalay,
  J.~Gupchup, J.~Klofas, L.~Xia, C.~Swarth, and S.~Matthews, ``{Turtle Nest
  Monitoring with Wireless Sensor Networks},'' in \emph{Proceedings of the
  American Geophysical Union, Fall Meeting}, 2007.

\bibitem{JBPR}
{National Estuarine Research Reserve}, ``{Jug Bay weather station (cbmjbwq)},''
  Available at \url{http://cdmo.baruch.sc.edu/QueryPages/anychart.cfm}.

\bibitem{DSI07}
J.~Gupchup, A.~Terzis, R.~Burns, and A.~Szalay, ``{Model-Based Event Detection
  in Wireless Sensor Networks},'' in \emph{Proceedings of the Workshop for Data
  Sharing and Interoperability on the World Wide Web (DSI 2007)}, Apr. 2007.

\end{thebibliography}
